\documentclass[journal,twoside,a4paper]{IEEEtran}

\usepackage{cite}      

\usepackage{graphicx}  

\usepackage{amsmath}   
\begin{document}
%
\title{Quantum Feedback Channels}
%
%
\author{{Garry~Bowen}
\thanks{G. Bowen is with the Centre for Quantum Computation, DAMTP, University of Cambridge, Wilberforce Road, Cambridge CB3 0WA, UK and the Department of Computer Science, University of Warwick, Coventry CV4 7AL, UK (e-mail: gab30@damtp.cam.ac.uk).}}%
%
%
%
\markboth{Submitted to IEEE Transactions on Information Theory}{Quantum Feedback Channels}
%




\newtheorem{theorem}{Theorem}
\newtheorem{lemma}{Lemma}
\newtheorem{defn}{Definition}
\newtheorem{corollary}{Corollary}

\maketitle

\begin{abstract}
In Shannon information theory the capacity of a memoryless communication channel cannot be increased by the use of feedback.  In quantum information theory the no-cloning theorem means that noiseless copying and feedback of quantum information cannot be achieved.  In this paper, quantum feedback is defined as the unlimited use of a noiseless quantum channel from receiver to sender.  Given such quantum feedback, it is shown to provide no increase in the entanglement--assisted capacities of a memoryless quantum channel, in direct analogy to the classical case.  It is also shown that in various cases of non-assisted capacities, feedback may increase the capacity of memoryless quantum channels.
\end{abstract}


%
\IEEEpeerreviewmaketitle

\section{Introduction}
%
%
%
%

\PARstart{I}{n} the Shannon theory of information transmission through noisy channels, the existence of feedback has been shown not to increase the capacity of a memoryless channel \cite{cover}.  A memoryless channel is defined as a noisy channel where the noise acts independently on each symbol sent through the channel.  The classical feedback channel is presumed to send an exact copy of the received symbol back to the sender, before the next symbol is sent through the noisy channel.  Even with this extra information the sender cannot increase the asymptotic rate at which information may be sent.
The use of feedback may also be extended to include any information sent from the receiver to the sender.  In this way, the feedback operation becomes a noiseless channel of arbitrary capacity from the receiver to the sender.  The capacity of a memoryless channel augmented by a noiseless feedback channel remains unchanged compared to the memoryless channel without augmentation by the feedback channel.

In the broader context of quantum information theory, the exact nature of feedback is a more difficult concept.  This is due to the quantum no-cloning theorem \cite{wootters82}.  The no-cloning theorem does not allow exact copying of unknown quantum states to take place.  Any such attempted copying process necessarily detracts from the average fidelity of the received states.
The process of feedback in the quantum scenario would therefore only be useful if it can be shown how \textit{any quantum information} passed from the receiver to the sender will affect the channel capacities.  Quantum feedback is thus defined as a noiseless quantum channel, of arbitrary capacity, from receiver to sender.  The receiver may process the received quantum states in any way, and send an arbitrary amount of quantum or classical information through the feedback channel.  In the case of cloning the state, the receiver is, of course, bounded in the fidelity with which this may be achieved \cite{buzek96,bruss98,gisin97}.

The various classical and quantum capacities of the channel assisted by classical feedback only, may also be considered.  However, for the entanglement--assisted capacities there is an equivalence between the quantum and classical feedback channels.  In this situation a noiseless classical feedback channel automatically becomes a noiseless quantum channel.  This is due to the availability of an unlimited supply of maximally entangled states and the process of quantum teleportation \cite{bennett93}.

It is known that pre-existing shared quantum entanglement, between the sender and receiver, may be used to increase the classical and quantum information capacities of memoryless quantum channels.  The closed expression for the entanglement assisted capacity is also known to mirror the expression for the classical information capacity through a noisy classical channel \cite{bennett01a}.  By utilizing quantum teleportation and quantum dense coding \cite{bennett92}, it may be shown that the entanglement assisted quantum capacity $Q_E$, is precisely half the entanglement--assisted classical capacity.

In this paper, it is shown that the entanglement--assisted capacity of a memoryless quantum channel cannot be increased by feedback.  Both the entanglement assisted capacity with classical feedback $C^{\, \mathrm{FB}}_E$, and the classical capacity with quantum feedback $C^{\, \mathrm{QFB}}$ (with or without prior shared entanglement), are equivalent to the entanglement--assisted capacity without feedback, that is,
\begin{equation}
C^{\, \mathrm{FB}}_E = C^{\, \mathrm{QFB}} = C_E
\end{equation}
for any memoryless quantum communication channel.  As the entanglement--assisted quantum capacities are exactly half the corresponding entanglement--assisted classical capacities, these are also unchanged by the use of feedback.  Furthermore, examples of memoryless channels for which the capacities with quantum or classical feedback exceed the unassisted quantum capacity are also given.  For quantum information the analogy with the Shannon theory is thus only shown to hold in the case of the entanglement--assisted capacities.

\section{Preliminaries}

The von Neumann entropy of a quantum state $\rho$ is defined by $S(\rho) = - \mathrm{Tr}\, \rho \log \rho$, with $\mathrm{Tr}$ denoting the trace operation.  For finite state quantum systems, this is equivalent to the Shannon entropy of the eigenvalues of $\rho$ \cite{nielsen}.  The quantum conditional entropy and quantum mutual information may be defined analogously as $S(A|B) = S(\rho_{AB}) - S(\rho_B)$, and $S(A:B) = S(\rho_A)+S(\rho_B)-S(\rho_{AB})$, respectively, for a bipartite quantum state $\rho_{AB}$.  The quantum mutual information provides an upper bound on the mutual information that may be generated between local measurement outcomes on the bipartite state \cite{kholevo73,holevo98,schumacher97,cerf98}.

The conditional quantum mutual information may be expressed in a number of equivalent forms,
\begin{align}
S(A:B|C) &= S(\rho_{AC}) + S(\rho_{BC}) - S(\rho_C) - S(\rho_{ABC}) \\
&= S(A:BC) - S(A:C) \label{eqn:vncmi_change}
\end{align}
for any tripartite state $\rho_{ABC}$.  From (\ref{eqn:vncmi_change}) it may be seen that this quantity represents the change in the quantum mutual information when one party gains access to an additional part of the quantum system.

The \textit{entanglement--assisted classical capacity} for a memoryless quantum channel $\Lambda$ is given by the maximum quantum mutual information that may be generated between the sender and receiver through the channel \cite{bennett01a},
\begin{equation}
C_E = \max_{\rho} \Big[ S(\rho) + S(\Lambda \rho) - S\big( (\mathcal{I} \otimes \Lambda ) |\Psi \rangle \langle \Psi| \big) \Big]
\label{eqn:C_E}
\end{equation}
where $|\Psi\rangle$ is a purification of the quantum state $\rho$, and $\mathcal{I}$ the identity map.


\section{Quantum Feedback and Entanglement--Assisted Capacities}

In this section, the classical capacity of a quantum channel assisted by a quantum feedback channel is shown to be no larger than the entanglement--assisted capacity of the channel.  This implies that the addition of quantum feedback cannot increase the entanglement--assisted capacity of the channel.

\subsection{The Feedback Capacity in Terms of Shared Ensembles}

Modeling an arbitrary quantum feedback protocol over many uses of a quantum channel may initially seem to present a difficult task.  However, by utilizing recursive arguments it can be shown that an arbitrary protocol may be bounded by the conditional mutual information generated by a single use of the channel.  In determining the classical capacity of a quantum channel it is helpful to introduce a class of ensembles that allow for simple analysis of the mutual information generated through the channel.

\begin{defn}
An $n$-ensemble of states is a set of density operators on $\mathcal{H}_M \otimes \mathcal{H}_Q^{\otimes n}$ with probabilities $p_i$, and an average state,
\begin{equation}
\rho_{MQ^{(n)}} = \sum_i p_i \, |i\rangle\langle i|_M \otimes \rho^i_{Q^{(n)}}
\label{eqn:n_ensemble}
\end{equation}
where the states $|i\rangle_M$ are orthogonal.  The notation $Q^{(n)}$ denotes the state acting on the $n$-fold tensor product space $\mathcal{H}_{Q_1}\otimes ... \otimes \mathcal{H}_{Q_n}$ and $Q_k$ denotes the state on $\mathcal{H}_{Q_k}$.
\end{defn}

The message space $M$ exists as a copy of the sample space from which Alice sends a message to Bob.  The index $n$ on the generated message $M$ has been suppressed as this should be clear from the context.  The message space is assumed to be generated from a finite alphabet stochastic process $\mathcal{M}$ that satisfies the asymptotic equipartition theorem \cite{cover}.  The states $|i\rangle_M$ are chosen to be orthogonal, as they may then be measured and copied arbitrarily, as for any normal classical message.  Each message state in $M$ has an associated quantum state in $Q^{(n)}$, which is viewed as the quantum encoding of the message in $M$.  The message generated by the source $\mathcal{M}$ for Alice to send is assumed not to change during transmission to Bob.  Alice is free, however, to change the encoding by changing the quantum state in $Q^{(n)}$.

To bound the capacity of the channel assisted by quantum feedback we rely on the bound for $I(M:Q^{(n)})$, the accessible mutual information, derived by Holevo \cite{kholevo73}.  The accessible mutual information is the mutual information of the random variables generated from the outcomes of any local measurements on the systems $M$ and $Q^{(n)}$.  The theorem is stated here without proof.
\begin{theorem}[Holevo]
For any $n$-ensemble of states the maximum accessible mutual information between $M$ and $Q^{(n)}$ is bounded above by,
\begin{equation}
I(M:Q^{(n)}) \leq S(M:Q^{(n)})
\label{eqn:holevo}
\end{equation}
with equality only when the ensemble states mutually commute.
\label{thm:holevo}
\end{theorem}

Combining (\ref{eqn:holevo}) with the following Lemma shows that the accessible mutual information generated through a memoryless quantum channel by an arbitrary quantum feedback protocol has an additive upper bound.
\begin{lemma}
For a memoryless quantum channel the maximum quantum mutual information generated for $n$ steps of a quantum feedback protocol is bounded by $n$ times the maximum quantum conditional mutual information that can be generated by any ensemble partially transmitted through the channel.  Explicitly,
\begin{equation}
S(M:Q^{(n)}Y^{(n)}) \leq n \max S(M:A|B)
\label{eqn:cond_bound}
\end{equation}
for the maximization over states $\rho_{MAB} = \sum_i p_i \, |i\rangle\langle i|_M \otimes \big(\Lambda_A \otimes\mathcal{I}_B\big)\rho^i_{AB}$ with $\dim \mathcal{H}_A = \dim \mathcal{H}_Q$, and $\Lambda$ the quantum channel.
\label{lem:max_change}
\end{lemma}
\begin{proof}
Any quantum feedback operation may be expressed in the following form:
\begin{itemize}
\item[i.]{Bob operates on the received state $Q_{k}$, previously received states $Q^{(k-1)}$, and a bipartite state $X_{k}Y^{(k)}$, with a unitary operation $U_{Q^{(k)}X_{k}Y^{(k)}}$.}
\item[ii.]{The state $X_{k}$ is transmitted through the noiseless feedback channel to Alice.}
\item[iii.]{Alice acts with a unitary operation $V^i_{Q_{k+1}X^{(k)}Z^{(k)}}$ on $X^{(k)}$, the ancilla states $Z^{(k)}$, and the next message state $Q_{k+1}$.}
\item[iv.]{Alice transmits state $Q_{k+1}$ through the channel to Bob.}
\end{itemize}
Note that all the operations on states $X$, $Y$ and $Z$, are inclusive of previous steps in the protocol.  The extension to a unitary process covers all possible quantum operations, as part of the ancilla states may be considered as an environment which is subsequently discarded.  Measurements followed by operations conditional on the outcome may be incorporated in a unitary description by recording the ``measurement'' outcome in an ancilla and using a conditional operation based on the value of the ancilla.  This method automatically includes averages over all possible outcomes of the measurements.

The most important fact of any such quantum feedback is that it cannot increase the quantum mutual information between the message state $M$ and the previously transmitted states.  Formally, this follows from the monotonicity of the quantum mutual information under trace preserving operations, where,
\begin{align}
S(M:Q^{(k)}Y^{(k)}) &\leq S(M:Q^{(k)}X_{k}Y^{(k)}) \\
&= S(M:Q^{(k)}Y^{(k-1)})
\end{align}
It is important to note that the total quantum mutual information shared by Alice and Bob \textit{can} increase, simply by sharing correlated states (including entangled states).  However, the message state is invariant under the feedback operations, as Alice is transmitting a preselected message.  Hence, the quantum mutual information between Alice's message and the transmitted states \textit{cannot} increase.  This allows the recursive removal of feedback operations to bound the expansion in terms of the pre-feedback states.  For $k=n$ we have,
\begin{align}
S(M:Q^{(n)}Y^{(n)}) &\leq S(M:Q^{(n)}Y^{(n-1)}) \\
&= S(M:Q_n|Q^{(n-1)}Y^{(n-1)}) \nonumber \\
&\phantom{=}\:+ S(M:Q^{(n-1)}Y^{(n-1)})
\label{eqn:feed_strip}
\end{align}
where the first term in (\ref{eqn:feed_strip}) is the quantum conditional mutual information of the state $\rho_{MQ^{(n)}Y^{(n-1)}} = \sum_i p_i \, |i\rangle\langle i|_M \otimes (\Lambda_{Q_n} \otimes \mathcal{I}_{Q^{(n-1)}Y^{(n-1)}})\rho^i_{Q^{(n)}Y^{(n-1)}}$.  By a recursive application on the second term of (\ref{eqn:feed_strip}), where $k = n-1$, it can be seen that the total quantum mutual information is bounded by the sum over the conditional quantum mutual information,
\begin{equation}
S(M:Q^{(n)}Y^{(n)}) \leq \sum_{k=1}^n S(M:Q_k|Q^{(k-1)}Y^{(k-1)})
\end{equation}
of each of the states $\rho_{MQ^{(k)}Y^{(k-1)}} = \sum_i p_i \, |i\rangle\langle i|_M \otimes (\Lambda_{Q_k} \otimes \mathcal{I}_{Q^{(k-1)}Y^{(k-1)}})\rho^i_{Q^{(k)}Y^{(k-1)}}$, for $1\leq k \leq n$.  The conditional quantum mutual information of each of these states is then necessarily less than the maximum over all such states, giving the inequality in (\ref{eqn:cond_bound}).
\end{proof}

\begin{theorem}
The capacity of a memoryless quantum channel $\Lambda$ utilizing quantum feedback is bounded above by the maximum conditional quantum mutual information that may be generated through a single use of the channel,
\begin{equation}
C^{\, \mathrm{QFB}} \leq \max S(M:A|B)
\label{eqn:QFB_bound}
\end{equation}
with the maximization over states of the form $\rho_{MAB} = \sum_i p_i \, |i\rangle\langle i|_M \otimes \big(\Lambda_A \otimes\mathcal{I}_B\big)\rho^i_{AB}$.
\end{theorem}
\begin{proof}
Following measurement of the output states and decoding, the process $M \rightarrow Q^{(n)}Y^{(n)} \rightarrow \tilde{M}$ forms a Markov chain, where $\tilde{M}$ is the decoded output message.  Here all the variables $M$, $Q^{(n)}Y^{(n)}$ and $\tilde{M}$ are classical random variables determined by the outcomes of measurements on the respective states.  The error probability for a code is given by $P_e^{(n)} = \mathrm{Prob}[M \neq \tilde{M}]$.  It is sufficient to show that an asymptotically vanishing error probability $P^{(n)}_e \rightarrow 0$ implies the entropy of the message source $\mathcal{M}$ cannot be greater than the right hand side of (\ref{eqn:QFB_bound}).  Thus,
\begin{align}
H(\mathcal{M}) &\leq \frac{H(M)}{n} \label{eqn:line1} \\
&= \frac{1}{n}\big[ H(M|\tilde{M}) + I(M:\tilde{M})\big] \label{eqn:line2} \\
&\leq \frac{1}{n}\Big[ 1 + nP_e^{(n)}\log |\mathcal{M}| + I(M:\tilde{M}) \Big] \label{eqn:line3} \\
&\leq \frac{1}{n}\Big[ 1 + nP_e^{(n)}\log |\mathcal{M}| + I(M:Q^{(n)}Y^{(n)}) \label{eqn:line4} \Big] \\
&\leq \frac{1}{n} + P_e^{(n)}\log |\mathcal{M}| + \max S(M:A|B) \label{eqn:line5}
\end{align}
where (\ref{eqn:line1}) follows from the definition of the entropy rate, (\ref{eqn:line3}) from Fano's inequality, (\ref{eqn:line4}) from the data processing inequality, and (\ref{eqn:line5}) from Theorem \ref{thm:holevo} and Lemma \ref{lem:max_change}.  Taking the limit as $n\rightarrow \infty$ implies $P_e^{(n)} \rightarrow 0$ and hence $H(\mathcal{M}) \leq \max S(M:A|B)$.
\end{proof}

The fact that the maximization over the quantity in (\ref{eqn:cond_bound}) is an additive bound has been utilized previously in the derivation of the unidirectional entanglement--assisted capacity of unitary operators acting on bipartite states \cite{bennett02}.

\subsection{Bounding the Quantum Feedback Capacity by $C_E$}

Firstly, it should be noted that the entanglement--assisted capacity augmented with classical feedback is equivalent to the capacity with quantum feedback $C^{\, \mathrm{FB}}_E = C^{\, \mathrm{QFB}}$.  The equivalence follows from the fact that classical feedback augmented with unlimited shared entanglement corresponds to a noiseless quantum feedback channel via teleportation \cite{bennett93}, and, a quantum feedback channel may be utilized to both share entanglement and communicate classical information.

Now, as a quantum feedback channel may be used to share arbitrarily large amounts of entanglement, the quantum feedback capacity for the channel is at least as great as the entanglement--assisted capacity without feedback $C^{\, \mathrm{QFB}} \geq C_E$.  For these capacities to be equal it need only be shown that the capacity with quantum feedback cannot exceed the entanglement--assisted capacity.
\begin{theorem}
The capacity of a memoryless quantum channel augmented by quantum feedback cannot exceed the entanglement--assisted capacity of the channel
\begin{equation}
C^{\, \mathrm{QFB}} \leq C_E
\label{eqn:feedback_bound}
\end{equation}
and similarly for the entanglement--assisted capacity augmented with classical feedback.
\end{theorem}
\begin{proof}
Given the ensemble,
\begin{equation}
\rho_{MAB} = \sum_i p_i \, |i\rangle\langle i|_M \otimes \rho^i_{AB}
\label{eqn:initial_ensemble2}
\end{equation}
the change in the quantum conditional mutual information is bounded by,
\begin{align}
\Delta &= S(M : A|B) \nonumber \\
&= S\big((\Lambda_A \otimes \mathcal{I}_B)\rho_{AB}\big) - S\big((\Lambda_A \otimes \mathcal{I}_{MB})\rho_{MAB}\big) \nonumber \\
&\phantom{=}\:- S(\rho_B) + S(\rho_{MB}) \nonumber \\
&\leq S\big((\Lambda_A \otimes \mathcal{I}_B)\rho_{AB}\big) - \sum_i p_i S\big((\Lambda_A \otimes \mathcal{I}_{B})\rho^i_{AB}\big) \nonumber \\
&\phantom{=}\:- S(\rho_B) + \sum_i p_i S(\rho^i_{B})
\label{eqn:mut_diff_bound} \\
&\leq S\Big( \Lambda_A \Big[ \sum_i p_i \rho^i_{A} \Big] \Big) + \sum_i p_i S\big( \rho^i_{B} \big) \nonumber \\
&\phantom{=}\:- \sum_i p_i S\Big( \big(\Lambda_A \otimes \mathcal{I}_B\big) \rho^i_{AB} \Big)
\label{eqn:expand_delta_chi}
\end{align}
where (\ref{eqn:mut_diff_bound}) follows from the concavity of the quantum conditional entropy, and (\ref{eqn:expand_delta_chi}) follows from the subadditivity of the von Neumann entropy $S(\omega_{AB}) \leq S(\omega_A) + S(\omega_B)$.

The remaining part of the proof follows the derivation of an upper bound on the entanglement assisted capacity given by Holevo \cite{holevo01}.  By the monotonicity of the conditional entropy $S(Q|R) = S(\rho_{QR}) - S(\rho_{R})$ \cite{lieb73}, each of the terms $-S(A'_i|B_i) = S\big( \rho^i_{B} \big) - S\big( (\Lambda_A \otimes \mathcal{I}_B) \rho^i_{AB} \big)$ satisfies the bound,
\begin{equation}
-S(A'_i|B_i) \leq -S(A'_i|B_iE_i) = -S(A'_i|R_i)
\label{eqn:ce_mono}
\end{equation}
where $\rho^i_{AR_i} = \rho^i_{ABE_i}$ is a purification of the state $\rho^i_{AB}$ with an environment $E_i$.  A pure state implies equality in the entropies of the reduced density matrices $S(\rho^i_A) = S(\rho^i_{R_i})$, and so from (\ref{eqn:ce_mono}) we have,
\begin{multline}
S\big( \rho^i_{B} \big) - S\big( (\Lambda_A \otimes \mathcal{I}_B) \rho^i_{AB} \big) \\
\leq S\big( \rho^i_{A} \big) - S\big( (\Lambda_A \otimes \mathcal{I}_{R_i}) \rho^i_{AR_i} \big) .
\end{multline}
Given a purification $\rho_{QR}$ of the state $\rho_Q$ with reference system $R$, the function $\rho_Q \rightarrow S(\rho_Q) - S\big( (\Lambda \otimes \mathcal{I}) \rho_{QR}\big)$ is concave in $\rho_Q$ \cite{holevo01}.  Hence, the last two terms of (\ref{eqn:expand_delta_chi}) satisfy the bound,
\begin{multline}
\sum_i p_i \Big[ S\big( \rho^i_{B} \big) - S\Big( \big(\Lambda_A \otimes \mathcal{I}_B\big) \rho^i_{AB} \Big) \Big] \\
\leq S(\rho_A) - S\big( (\Lambda \otimes \mathcal{I}) \rho_{AR}\big)
\label{eqn:con_con}
\end{multline}
where $\rho_{AR}$ is a purification of the state $\rho_A = \sum_i p_i \rho^i_A$.  Combining the bound in (\ref{eqn:con_con}) with (\ref{eqn:expand_delta_chi}) and maximizing over the ensemble implies,
\begin{equation}
C^{\, \mathrm{QFB}} \leq \max_{\rho_A} \Big[ S(\Lambda \rho_A) + S(\rho_A) - S\big( (\Lambda \otimes \mathcal{I}) \rho_{AR}\big) \Big]
\label{eqn:final_ineq}
\end{equation}
The right hand side of (\ref{eqn:final_ineq}) is equivalent to the expression for the entanglement--assisted capacity $C_E$ in (\ref{eqn:C_E}), thus demonstrating the required inequality in (\ref{eqn:feedback_bound}).
\end{proof}

\section{Feedback and Unassisted Capacities}

The invariance of the entanglement--assisted capacity with the addition of feedback does not extend to all unassisted capacities of the channel.  As Bob can share an unlimited number of maximally entangled states with Alice through a quantum feedback channel, the capacity with quantum feedback is equal to the entanglement assisted capacity.  Any channel with an entanglement--assisted capacity higher than the corresponding unassisted capacity therefore has a higher capacity with quantum feedback.  Two examples of such channels are the qubit erasure channel \cite{bennett97} and the qubit depolarizing channel with entanglement fidelity $0.25< F \leq 0.75$ \cite{bennett96}.

Whether or not classical feedback can increase the unassisted classical capacity of any noisy quantum channel is still not known, although some partial results have been obtained \cite{bowen03a}.  The classical capacities of memoryless quantum channels are therefore related by the inequalities,
\begin{equation}
C \leq C^{\, \mathrm{FB}} \leq C^{\, \mathrm{QFB}} = C^{\, \mathrm{FB}}_E = C_E
\label{eqn:class_caps}
\end{equation}
with at least one inequality in (\ref{eqn:class_caps}) being strict for channels with $C \neq C_E$.

In the case of the unassisted quantum capacity $Q$ of a channel, the capacity may be exceeded when the channel is augmented by classical feedback.  An example of a family of channels for which this is possible is the qubit erasure channel \cite{bennett97}.  The qubit erasure channel has a known quantum capacity of $Q_{\mathrm{erasure}} = 1 - 2\epsilon$, for erasure probability $\epsilon$.  To exceed the unassisted quantum capacity for the quantum erasure channel, Alice begins by sharing maximally entangled states through the channel, which arrive intact with probability $1-\epsilon$, and are rendered useless with probability $\epsilon$.  Bob informs Alice, via the feedback channel, whether or not the transmission has been successful.  In this way Alice and Bob can share maximally entangled states at a rate $R^{\mathrm{FB}}_E = 1 - \epsilon$.  On $N$ uses of the channel Alice and Bob need only utilize $M<N$ channels in order to share enough entanglement to use an entanglement--assisted quantum code \cite{bowen02b} for the remaining $N-M$ channels.  Thus $MR^{\mathrm{FB}}_E = (N-M)\mathcal{E}_Q$, for $\mathcal{E}_Q$ the minimum entanglement required for the entanglement--assisted quantum code.  The asymptotic rate for this protocol is then obtained by solving the equation $NQ^{\mathrm{FB}*} = (N-M)Q_E$, for which we find,
\begin{equation}
Q^{\mathrm{FB}*} = \left( \frac{R^{\mathrm{FB}}_E}{R^{\mathrm{FB}}_E + \mathcal{E}_Q} \right) Q_E .
\label{eqn:QFBlower}
\end{equation}
For the qubit erasure channel all the quantities on the right of (\ref{eqn:QFBlower}) are known, with $R^{\mathrm{FB}}_E = 1 - \epsilon$, $\mathcal{E}_Q = \epsilon$, and $Q_E = 1-\epsilon$, and thus,
\begin{equation}
Q^{\mathrm{FB}*}_{\mathrm{erasure}} = 1 - 2\epsilon + \epsilon^2 > Q_{\mathrm{erasure}}
\end{equation}
whenever $0 < \epsilon < 1$.  The qubit erasure channel is also an example of a channel for which $Q < Q_2$, where $Q_2$ is the quantum capacity with a two-way classical side channel.  It is known that forward classical communication provides no increase in the asymptotic rate for the unassisted quantum capacity, $Q = Q_{1 \rightarrow}$ \cite{bennett96,barnum00}.  It may well be the case that the addition of forward classical communication also provides no increase in the quantum capacity assisted by classical feedback, which would lead to the equality $Q^{\, \mathrm{FB}} = Q_2$.

\section{Conclusion}

In summary, it is known that the use of feedback in classical information theory cannot increase the capacity of a memoryless channel.  In this paper, it was shown that an analogous result will hold for information transmitted through memoryless quantum channels supplemented with quantum feedback, but \textit{only} in terms of the entanglement assisted capacities of the channel.  For channels without entanglement assistance the capacities with quantum feedback may be increased up to the corresponding entanglement--assisted capacities.  Furthermore, an example is given of a family of channels for which classical feedback increases the quantum capacity.


%
%


\section*{Acknowledgment}
The author would like to thank Sougato Bose for helpful discussions.


\begin{thebibliography}{10}
\providecommand{\url}[1]{#1}
\def\UrlFont{\rmfamily}
\providecommand{\newblock}{\relax}
\providecommand{\bibinfo}[2]{#2}
\providecommand\BIBentrySTDinterwordspacing{\spaceskip=0pt\relax}
\providecommand\BIBentryALTinterwordstretchfactor{4}
\providecommand\BIBentryALTinterwordspacing{\spaceskip=\fontdimen2\font plus
\BIBentryALTinterwordstretchfactor\fontdimen3\font minus
  \fontdimen4\font\relax}
\providecommand\BIBforeignlanguage[2]{{%
\expandafter\ifx\csname l@#1\endcsname\relax
\typeout{** WARNING: IEEEtran.bst: No hyphenation pattern has been}%
\typeout{** loaded for the language `#1'. Using the pattern for}%
\typeout{** the default language instead.}%
\else
\language=\csname l@#1\endcsname
\fi
#2}}

\bibitem{cover}
T.~M. Cover and J.~A. Thomas, \emph{Elements of Information Theory}.\hskip 1em
  plus 0.5em minus 0.4em\relax New York: Wiley, 1991.

\bibitem{wootters82}
W.~K. Wootters and W.~H. Zurek, ``A single quantum cannot be cloned,''
  \emph{Nature}, vol. 299, pp. 802--803, 1982.

\bibitem{buzek96}
V.~Bu\v{z}ek and M.~Hillery, ``Quantum copying: Beyond the no-cloning
  theorem,'' \emph{Phys. Rev. A}, vol.~54, pp. 1844--1852, 1996.

\bibitem{bruss98}
D.~Bru{\ss}, D.~P. DiVincenzo, A.~Ekert, C.~A. Fuchs, C.~Macchiavello, and
  J.~A. Smolin, ``Optimal universal and state-dependent quantum cloning,''
  \emph{Phys. Rev. A}, vol.~57, pp. 2368--2378, 1998.

\bibitem{gisin97}
N.~Gisin and S.~Massar, ``Optimal quantum cloning machines,'' \emph{Phys. Rev.
  Lett.}, vol.~79, pp. 2153--2156, 1997.

\bibitem{bennett93}
C.~H. Bennett, G.~Brassard, C.~Cr\'{e}peau, R.~Jozsa, A.~Peres, and W.~K.
  Wootters, ``Teleporting an unknown quantum state via dual classical and
  Einstein-Podolsky-Rosen channels,'' \emph{Phys. Rev. Lett.}, vol.~70, pp.
  1895--1899, 1993.

\bibitem{bennett01a}
C.~H. Bennett, P.~W. Shor, J.~A. Smolin, and A.~V. Thapliyal,
  ``Entanglement-assisted capacity of a quantum channel and the reverse Shannon
  theorem,'' \emph{IEEE Trans. Inform. Theory}, vol.~48, pp. 2637--2655, 2002.

\bibitem{bennett92}
C.~H. Bennett and S.~J. Wiesner, ``Communication via one- and two-particle
  operators on Einstein-Podolsky-Rosen states,'' \emph{Phys. Rev. Lett.},
  vol.~69, pp. 2881--2884, 1992.

\bibitem{nielsen}
M.~A. Nielsen and I.~L. Chuang, \emph{Quantum Computation and Quantum
  Information}.\hskip 1em plus 0.5em minus 0.4em\relax Cambridge: Cambridge
  University Press, 2000.

\bibitem{kholevo73}
A.~S. Holevo, ``Bounds for the quantity of information transmitted by a quantum
  communication channel,'' \emph{Probl. Peredachi Inf.}, vol.~9, pp. 3--11,
  1973.

\bibitem{holevo98}
------, ``The capacity of the quantum channel with general signal states,''
  \emph{IEEE Trans. Inform. Theory}, vol.~44, pp. 269--273, 1998.

\bibitem{schumacher97}
B.~Schumacher and M.~D. Westmoreland, ``Sending classical information via noisy
  quantum channels,'' \emph{Phys. Rev. A}, vol.~56, pp. 131--138, 1997.

\bibitem{cerf98}
N.~J. Cerf and C.~Adami, ``Information theory of quantum entanglement and
  measurement,'' \emph{Physica D}, vol. 120, pp. 62--81, 1998.

\bibitem{bennett02}
C.~H. Bennett, A.~Harrow, D.~W. Leung, and J.~A. Smolin, ``On the capacities of
  bipartite Hamiltonians and unitary gates,'' \emph{IEEE Trans. Inform.
  Theory}, vol.~49, pp. 1895--1911, 2003.

\bibitem{holevo01}
A.~S. Holevo, ``On entanglement-assisted classical capacity,'' \emph{J. Math.
  Phys.}, vol.~43, pp. 4326--4333, 2002.

\bibitem{lieb73}
E.~H. Lieb and M.~B. Ruskai, ``Proof of the strong subadditivity of
  quantum-mechanical entropy,'' \emph{J. Math. Phys.}, vol.~14, pp. 1938--1941,
  1973.

\bibitem{bennett97}
C.~H. Bennett, D.~P. DiVincenzo, and J.~A. Smolin, ``Capacities of quantum
  erasure channels,'' \emph{Phys. Rev. Lett.}, vol.~78, pp. 3217--3220, 1997.

\bibitem{bennett96}
C.~H. Bennett, D.~P. DiVincenzo, J.~A. Smolin, and W.~K. Wootters,
  ``Mixed-state entanglement and quantum error correction,'' \emph{Phys. Rev.
  A}, vol.~54, pp. 3824--3851, 1996.

\bibitem{bowen03a}
G.~Bowen and R.~Nagarajan, ``On feedback and the classical capacity of a noisy quantum channel", quant-ph/0305176.

\bibitem{bowen02b}
G.~Bowen, ``Entanglement required in achieving entanglement-assisted channel
  capacities,'' \emph{Phys. Rev. A}, vol.~66, p. 052313, 2002.

\bibitem{barnum00}
H.~Barnum, E.~Knill, and M.~A. Nielsen, ``On quantum fidelities and channel
  capacities,'' \emph{IEEE Trans. Inform. Theory}, vol.~46, pp. 1317--1329,
  2000.

\end{thebibliography}
\end{document}